\documentclass[12pt]{article}
\usepackage{epsf}
\usepackage{amsmath}
\usepackage{graphics}
\usepackage{cite}

\renewcommand{\thefootnote}{\#\arabic{footnote}}
\begin{document}

\newcommand{\gtrsim}{ \mathop{}_{\textstyle \sim}^{\textstyle >} }
\newcommand{\lesssim}{ \mathop{}_{\textstyle \sim}^{\textstyle <} }

\newcommand{\rem}[1]{{\bf #1}}

\renewcommand{\thefootnote}{\fnsymbol{footnote}}
\setcounter{footnote}{0}
\begin{titlepage}

\def\thefootnote{\fnsymbol{footnote}}

\begin{center}
\hfill April 2015\\
\vskip .5in
\bigskip
\bigskip
{\Large \bf Lepton Number Conservation, Long-lived Quarks and 
Superweak Bileptonic Decays}

\vskip .45in

{\bf Paul Howard Frampton}\footnote{e-mail address: paul.h.frampton@gmail.com}

{\em Oxford, UK.}

\end{center}

\vskip .4in
\begin{abstract}
In the upcoming LHC Run 2, at $\sqrt{s} \sim13$ TeV, it is suggested to seek unusually charged 
($Q= -4/3$ and $+5/3$) quarks with mass $M_Q \sim 3$ TeV
which carry lepton number ($L = +2$ and $-2$ respectively) and decay superweakly to a bilepton $Y$ with mass $M_Y\sim2.5$ TeV and a usual quark.
These long-lived decays will have displaced decay vertices and
produce a striking final state in $pp$ which contains two separated jets together with
two pairs of correlated like-sign charged leptons. Such a process was inaccessible energetically in LHC Run 1 with $\sqrt{s} \sim8$ TeV.  The simplest theoretical explanation is the 331-model which has new physics necessarily below $4$ TeV and which explains the existence of three families by anomaly cancellation. 
\end{abstract}
\end{titlepage}

\renewcommand{\thepage}{\arabic{page}}
\setcounter{page}{1}
\renewcommand{\thefootnote}{\#\arabic{footnote}}

\newpage

\section{Introduction}
\bigskip

\noindent
In addition to the theoretical predictions for the LHC
of supersymmetry and dark matter, the discovery of either of which would be revolutionary, 
it is worth being
more conservative and to consider instead the ancient art of model-building in 
gauge theories which extend the standard model and
are motivated and testable.
In particular, we here suggest that LHC
experimentalists seek unusually-charged quarks ($Q=-4/3, +5/3$) which
are produced strongly and decay slowly by weaker than weak interactions,
are constrained to lie below $4$ TeV, and motivated by an explanation of three
families.

\bigskip

\noindent
The 331 model\cite{F,PP}  has provoked sufficient interest that there exist
a number of studies of
its phenomenological ramifications. One aspect which has, however, escaped much
attention is the issue of lepton number ($L$) conservation and the role it plays in
suppressing the decay rate for the heavy quarks. Although 
there are reviews of 331 bilepton physics\cite{CR,CD}, the slow
decays of the 331 heavy quarks have not been previously emphasized.
The upgraded LHC seems tailor-made for discovery
of these heavy quarks and its Run 2 could expose them.

\bigskip

\noindent
The familiar quarks $(u, d, c, s, t, b)$ have baryon number $B=1/3$ and $L=0$.
The familiar leptons $(e-, \mu-, \tau-, \nu_e, \nu_{\mu}, \nu_{\tau})$ have
$B=0$ and $L=1$. The exotically-charged quarks of the 331 model carry nonzero
$L$ as follows: $D$ and $S$ have $B=1/3$ and $L=2$; $T$ has $B=1/3$ and
$L=-2$.

\bigskip

\noindent
An important final-state particle at the LHC is the penetrating and unstable
muon which decays via the weak interaction
\begin{equation}
\mu^- \rightarrow e^- + \bar{\nu}_e + \nu_{\mu}
\label{mudecay}
\end{equation}
with a long lifetime $\tau_{\mu} \sim 2\times10^{-6}$s according \cite{vRS1,vRS2} to the tree-level formula
\begin{equation}
\tau_{\mu} = \frac{g_2^4 M_{\mu}^5}{12288 \pi M_W^4}
\label{mulifetime}
\end{equation} 
where $g_2$ is the electroweak $SU(2)$ gauge coupling with $g_2^2 = 8M_W^2G_F$.
Other than the muon, the only long-lived charged particles in the standard model
are the stable electron and proton. But there may be about to appear an entirely 
new breed of metastable charged elementary particles to enter this small group. 

\bigskip

\noindent
It is well-known and investigated that if there exists 
a fourth family of quarks,
then they can mix only very little with the first three families
because the $3\times3$ CKM matrix \cite{C,KM}
is close to being unitary. Hence, the additional quarks
of the fourth family would have interesting 
long lifetimes as discussed  in \cite{FH,MRSY}.
In the 331-model it is assumed
that such sequential quarks do not exist and it is predicted that there
are only three families.

\bigskip

\noindent
Here we are interested in quarks which are long-lived for a different reason,
namely $L$ conservation. The consequent superweak interaction
is mediated by $Y$ bilepton intermediate vector bosons
in the 331-model and a possible
particle discovery at LHC is of a sibling to the $W^{\pm}$.
For instance $Y^-$ mediates the abnormal muon decay
\begin{equation}
\mu^- \rightarrow e^- + {\nu}_e + \bar{\nu}_{\mu}
\label{muotherdecay}
\end{equation}
which is suppressed relative to the normal decay, Eqs. (\ref{mudecay},\ref{mulifetime})
by a factor 
\begin{equation}
f= (M_W/M_Y)^4. 
\label{f}
\end{equation}
and, given $M_W \simeq 80$ GeV, the question is whether the value of $M_Y$ 
can be arbitrarily large. In the present context the answer is no.

\bigskip

\noindent
In the
331-model there is an important theoretical upper limit $M_Y \leq 4TeV$  for the
symmetry breaking to the standard model, arising from the renormalization
group behavior of the electroweak mixing angle and the group
embedding. The value $\sin^2 \theta (M_Z) = 0.231$ runs upward with
energy scale and reaches $\sin^2 \theta (E) = 0.250$, a singular point
of the embedding $SU(2)_L \subset SU(3)_L$, at $E=4$ TeV.
This was first analysed in \cite{F} and has been much
more recently confirmed in \cite{Buras}. This intrinsic 331 upper limit is
what underlies the claim that the new physics is at a mass scale especially
befitting LHC's Run 2.

\bigskip

\noindent
Theoretically then $M_Y \lesssim 4$ TeV while experimentally
$M_Y \geq 1.5$ TeV. We may reasonably take
$M_Y=2.5$ TeV ($\simeq 10\sqrt{10} M_W$) as an illustration
whereupon the suppression factor $f$ in Eq. (\ref{f}) is $f \simeq10^{-6}$.
The experimental upper limit for process Eq.(\ref{muotherdecay})
 is \cite{PDG} disappointing, the branching ratio being restricted
 merely to $\leq 1.2\%$. The 331
 prediction is that this
 branching ratio is four orders of magnitude smaller, $\sim 10^{-6}$.

\bigskip

\noindent
 By superweak
interaction we therefore mean the
weak interaction further suppressed for bilepton mediation
by the factor $f$ in Eq.(\ref{f}) relative to the $W$ exchange. Superweakness
implies that the exotic quarks $(D, S, T)$ 
are long-lived.

\newpage

\bigskip

\section{Long-Lived Quarks}

\bigskip

\noindent
In the 331-model which requires exactly
three families there are three additional exotic quarks($D, S, T$), one in each family. The
gauge group is $SU(3)_C \times SU(3)_L \times U(1)$ and for the first family
the quarks are in the triplet and three singlets of $SU(3)_L$
\begin{equation}
\left( \begin{array}{c} u^{\alpha} \\
d^{\alpha} \\
D^{\alpha} 
\end{array}
\right)_L        ~~~ \bar{D}_{L.\alpha}, ~~~ \bar{d}_{L.\alpha}, ~~~ \bar{u}_{L,\alpha},
\end{equation}
and similarly for the second family
\begin{equation}
\left( \begin{array}{c} c^{\alpha} \\
s^{\alpha} \\
S^{\alpha} 
\end{array}
\right)_L        ~~~ \bar{S}_{L.\alpha}, ~~~ \bar{s}_{L.\alpha}, ~~~ \bar{c}_{L,\alpha}.
\end{equation}
The quarks of the third family are assigned differently, in one antitriplet and three singlets
\begin{equation}
\left( \begin{array}{c} T^{\alpha} \\
t^{\alpha} \\
b^{\alpha} 
\end{array}
\right)_L        ~~~ \bar{b}_{L.\alpha}, ~~~ \bar{t}_{L.\alpha}, ~~~ \bar{T}_{L,\alpha}.
\end{equation}

\bigskip

\noindent
The established weak gauge bosons $(W^-, W^0, W^-)$ with 
$W^0 \equiv Zcos\theta + \gamma\sin\theta$
are augmented by five more, a $Z^{'}$ and four bileptons 
$(Y^+, Y^{++})$ ($L=-2$) and $(Y^-, Y^{--})$ ($L=+2$).

\bigskip

\noindent
The superweak decays of $D$ are (we exhibit only the muonic decays,
the most readily detected)
\begin{equation}
D \rightarrow u + Y^{--} \rightarrow u + \mu^- + \mu^-
\label{Ddecay}
\end{equation}
has a displaced vertex in the silicon detector. There is the alternative equally long-lived
decay
\begin{equation}
D \rightarrow d + Y^{-} \rightarrow d + \mu^- + \nu_{\mu},
\label{Ddecay2}
\end{equation}
but the neutrino $\nu_{\mu}$ makes process Eq. (\ref{Ddecay2}) far more challenging to detect 
than Eq.(\ref{Ddecay}).

\bigskip

\noindent
The sequential second family exotic quark $S$ has similar long-lived decays
\begin{eqnarray}
S &\rightarrow& c + Y^{--} \rightarrow c + \mu^- + \mu^- \nonumber \\
S &\rightarrow& s + Y^{-} \rightarrow s + \mu^- + \nu_{\mu}
\label{Sdecay}
\end{eqnarray}

\bigskip

\noindent
In the 331 model the third family, on the other hand, the $T$ has $Q=+4/3$ and lepton number $L=-2$
so that its long-lived decays have flipped electric charges
\begin{eqnarray}
T &\rightarrow& b + Y^{++} \rightarrow b + \mu^+ + \mu^+ \nonumber \\
T &\rightarrow& t + Y^{+} \rightarrow t + \mu^+ + \bar{\nu}_{\mu}
\label{Tdecay}
\end{eqnarray}

\bigskip

\noindent
The antiquarks ($\bar{D}, \bar{S}, \bar{T}$) have superweak decays into the corresponding charge conjugate final states
\begin{eqnarray}
\bar{D} &\rightarrow& \bar{u} + Y^{++} \rightarrow \bar{u} + \mu^+ + \mu^+ \nonumber \\
\bar{D} &\rightarrow& \bar{d} + Y^+ \rightarrow \bar{d} + \mu^+ + \nu_{\mu} \nonumber \\
\bar{S} &\rightarrow& \bar{c} + Y^{++} \rightarrow \bar{c} + \mu^+ + \mu^+ \nonumber \\
\bar{S} &\rightarrow& \bar{s} + Y^+ \rightarrow \bar{s} + \mu^+ + \nu_{\mu} \nonumber \\
\bar{T} &\rightarrow& \bar{b} + Y^{--} \rightarrow \bar{b} + \mu^- + \mu^- \nonumber \\
\bar{T} &\rightarrow& \bar{t} + Y^{-} \rightarrow \bar{t} + \mu^- + \bar{\nu}_{\mu}
\end{eqnarray}

\bigskip

\noindent
We take $M_Y=2.5$ TeV and can assume a normal quark mass hierarchy
is $M_T > M_S > M_D$ with $M_D \sim 3$ TeV. The lowest threshold will then
be for $\bar{D}D$ pair production by strong interactions so the most interesting
event would be $pp \rightarrow \bar{D}D$ + any.

\bigskip

\noindent
By strong interactions, two gluons can produce the $\bar{D}D$ pair, practicable only at Run 2 of the LHC with $13 $ TeV. The $\bar{D}, D$ quarks
being long-lived will travel a macroscopic distance from the production vertex.

\bigskip

\noindent
The $\bar{D},D$ quarks decay to bileptons and the most striking signature would
surely be an event with
\begin{equation}
\bar{D} \rightarrow \bar{u} + Y^{++} \rightarrow \bar{u} + \mu^+ + \mu^+ 
\label{barD}
\end{equation}
and the counterpart
\begin{equation}
D \rightarrow u + Y^{--} \rightarrow u + \mu^- + \mu^-
\label{D}
\end{equation}

\bigskip

\noindent
In Eqs.(\ref{barD},\ref{D}), the light quarks $\bar{u}, u$ will hadronize to high-energy
jets which
the LHC physicists are well equipped to reconstruct. The two jets will originate from the
separate displaced decay vertices for $\bar{D}$ and $D$. The like-sign pairs
of muons will centre around an invariant bilepton mass $M_Y \sim 2.5$ TeV, which
can be determined. 

\bigskip

\noindent
This bilepton mass can subsequently be confirmed by the other channels
$\mu^-\nu_{\mu}$, $e^-e^-$, $e^-\nu_e$, $\tau^-\tau^-$, etc.

\newpage

\section{Discussion}

\bigskip

\noindent
The two most heralded targets for the LHC, beyond the Higgs boson, were
to confirm weak-scale supersymmetry and to produce dark matter.
If weak-scale supersymmetry existed, it was expected to appear
in the 2009-2013 Run 1 at $\sim 8$ TeV, but did not. The excluded
parameters narrow the likelihood of its
discovery in Run 2. Once weak-scale supersymmetry is abandoned,
the link between the weak scale and dark matter mass is lost. The masses for 
suggested DM candidates range from axions with mass $\sim 10^{-15}$ GeV 
to black holes with mass $\sim 10^{62}$ GeV so it would now require 
remarkably good fortune for it to show up at the LHC.
The possibility of extra spatial dimensions large enough to be detected
at the LHC is not strongly motivated.

\bigskip

\noindent
There are not many theoretical models with a strong
reason to expect the relevant new
physics scale to be specifically in the LHC Run 2 ($13$ TeV) regime, as opposed
to the Run 1 ($8$ TeV) one.  Among these, the 331 model does naturally contain an
multi-TeV scale \cite{F,Buras} in its analysis. Its long-lived charged
quarks are predicted in the appropriate mass range hence more likely to be 
produced in Run 2 than Run 1.
The signature of such an event is striking and although this would not immediately
explain all the parameters of the standard model it will give a second
stronger explanation of why
there exist three families beyond the ingenious observation in \cite{KM} that it
accommodates the observed CP violation in flavor-changing weak interactions.

\bigskip

\noindent
If such a discovery is made, what is the next step in the theory? 
It would suggest even further
cousins of the $W$ and $Y$ gauge bosons which could appear in
additional $SU(3)$ factors. Nonabelian subsumption of the $U(1)_Y$ gauge group factor is 
hinted at by avoidance
of the Landau pole. At present this remains idle speculation until an
electroweak $SU(3)$ has empirical evidence which would, nevertheless, firmly justify the
construction of higher energy apparatus to answer further questions.

\end{document}